\documentclass[obeyspaces,
  journal=pasa,
  manuscript=research-paper,
  year=2025,
  volume=YY,
]{cup-journal}

\usepackage{graphicx} 
\usepackage{amsmath}
\usepackage{amssymb,microtype,siunitx,booktabs}
\usepackage[skip=0.5ex]{subcaption}
\usepackage{url}
\usepackage{hyperref} 

\newcommand{\degr}{$^\circ$}
\newcommand{\arcmin}{$'$}
\newcommand{\arcsec}{$''$}
\newcommand{\zph}{$z_{\rm phot}$}
\newcommand{\zsp}{$z_{\rm spec}$}

\newcommand{\Msun}{~M$_{\odot}$}

\title{Two intersecting radio shells: relics of galaxy merger shocks\,?}

\author{B\"arbel S. Koribalski}
\affiliation{Australia Telescope National Facility, CSIRO, Space and Astronomy, P.O. Box 76, Epping, NSW 1710, Australia}
\alsoaffiliation{Western Sydney University, Locked Bag 1797, Penrith South DC, NSW 2751, Australia}
\email[B.S. Koribalski]{Baerbel.Koribalski@csiro.au}
\author{Klaus Dolag}
\affiliation{Universit\"ats-Sternwarte, Fakult\"at f\"ur Physik, Ludwig-Maximilians-Universit\"at M\"unchen, Scheinerstr.~1, D-81679 M\"unchen, Germany}
\alsoaffiliation{Max-Planck-Institut f\"ur Astrophysik, Karl-Schwarzschildstr. 1, D-85748, Garching, Germany} 
\author{Ildar Khabibullin}
\affiliation{Universit\"ats-Sternwarte, Fakult\"at f\"ur Physik, Ludwig-Maximilians-Universit\"at M\"unchen, Scheinerstr.~1, D-81679 M\"unchen, Germany}
\alsoaffiliation{Max-Planck-Institut f\"ur Astrophysik, Karl-Schwarzschildstr. 1, D-85748, Garching, Germany} 
\author{Ludwig M. B\"oss}
\affiliation{Department of Astronomy and Astrophysics, The University of Chicago, William Eckhart Research Center, 5640 S. Ellis Ave. Chicago, IL 60637}
\author{Anna Ivleva}
\affiliation{Universit\"ats-Sternwarte, Fakult\"at f\"ur Physik, Ludwig-Maximilians-Universit\"at M\"unchen, Scheinerstr.~1, D-81679 M\"unchen, Germany}
\author{Ray P. Norris}
\affiliation{Australia Telescope National Facility, CSIRO, Space and Astronomy, P.O. Box 76, Epping, NSW 1710, Australia}
\alsoaffiliation{Western Sydney University, Locked Bag 1797, Penrith South DC, NSW 2751, Australia}

\received {15 July 2025}
\revised  {dd mm 2025}
\accepted {dd mm 2025}
\published{dd mm 2025}


\begin{document}

\begin{abstract}
We present the discovery of two intersecting radio shells, likely radio relics, surrounding a compact galaxy group dominated by a massive elliptical galaxy. The shells present as partial, edge-brightened rings with diameters of $\sim$240\arcsec\ ($\sim$720~kpc) each and resemble a pair of odd radio circles. The central galaxy, WISEA J184105.19--654753.8, which shows signs of interactions, is radio bright, has a stellar mass of $3.1 \times 10^{11}$\Msun\ (for a redshift of \zph\ $\sim 0.18$) and is located in the intersect region. The double radio shell system, which we refer to as ORC~J1841--6547 (also known as ORC~6), was detected in 944~MHz radio continuum images obtained with Phased Array Feeds on the Australian Square Kilometre Array Pathfinder (ASKAP). The more prominent, north-western shell may be associated with an X-ray detection, while the weaker, south-eastern shell has no counterpart at non-radio wavelength. We propose outwards moving shocks from galaxy mergers driving into the intragroup medium, re-energising relic radio lobes, as a possible formation scenario for the observed radio shells. We conclude that at least some ORCs are shock-energised relics in the outskirts of galaxy groups, which originate during the merger evolution of the brightest group galaxy.
\end{abstract}

\section{Introduction} 
\label{sec:intro}

Recent discoveries of large-scale diffuse radio sources with the Australian Square Kilometre Array Pathfinder (ASKAP), such as cluster relics and halos \citep{Brueggen2021, Venturi2022, Riseley2022, Duchesne2024, Koribalski-Veronica2024}, extended radio galaxies \citep[e.g.,][]{Gurkan2022, Velovic2022, Venturi2022, Koribalski2024-Corkscrew, Koribalski2025}, odd radio circles \citep{Norris2021, Koribalski2021} and an intergalactic supernova remnant \citep{Filipovic2022}, give just a glimpse of what to expect from the on-going sky surveys \citep[e.g.,][]{EMU, EMU-PS, Koribalski2012, Koribalski2020}. Through a combination of wide-field imaging, good resolution and sensitivity, ASKAP radio surveys covering frequencies from 700 to 1800~MHz have become a treasure trove of findings \citep[see highlights in][]{Koribalski2022}, continuing to surprise and inspire new techniques to explore the vast volumes of data \citep{Gupta2022, Gupta2023, Segal2023}. Among the listed sources, odd radio circles (ORCs) stand out as one of the most peculiar and interesting. \\

The first ORC was discovered by \citet{Norris2021} in ASKAP 944~MHz radio continuum data from the `Evolutionary Map of the Universe' Pilot Survey \citep[EMU-PS,][]{EMU-PS}. They also found a peculiar ORC pair which is now considered to have a different origin, namely the bent lobes of a re-started radio galaxy (Macgregor et al., in prep.). The second single ORC, also discovered by \citet{Norris2021}, was detected in 325~MHz radio continuum data from the Giant Metrewave Radio Telescope (GMRT). The discovery of a third single ORC with ASKAP by \citet{Koribalski2021} highlighted that the massive elliptical galaxies at their centres are likely responsible for their formation. Two further ORCs were recently found in MeerKAT radio continuum images: ORC J1027--4422 \citep{Koribalski-Veronica2024} at relatively low Galactic latitude, making it difficult to study the likely host galaxy and its environment, and ORC J0219--0505 \citep{Norris2025}. This means, five bonafide single ORCs are currently known, undetected at non-radio wavelengths, and each centred on a massive elliptical galaxy with redshifts $z \sim 0.27 - 0.55$ and stellar masses $\gtrsim$10$^{11}$\Msun. The diameters of these ORCs are $\sim$1\arcmin, corresponding to $\sim$300 -- 500~kpc at the above redshifts. 

ORCs somewhat resemble (but are smaller, fainter, and more ring-like than) double radio relics in the outskirts of merging galaxy clusters \citep[e.g.,][]{Bagchi2006, vanWeeren2019, Koribalski-Veronica2024}. Using high-resolution cosmological simulations, \citet{Dolag2023} find that ORCs occasionally result from outwards moving merger shocks during the evolution of the growing central elliptical galaxy in groups. This scenario can also explain the formation of two nearby radio shell systems recently found in ASKAP images --- Physalis \citep[$z = 0.017$,][]{Koribalski2024-Physalis}, and Cloverleaf \citep[$z = 0.046$,][Koribalski et al., in prep.]{Dolag2023} --- suggesting that ORCs are radio relics around galaxy groups, formed through merger shocks expanding into remnant plasma from ageing radio lobes \citep[][]{Koribalski2024-Physalis}.  \\

\vspace{0.2cm}

\begin{figure*}[ht] 
\centering
  \includegraphics[width=17cm]{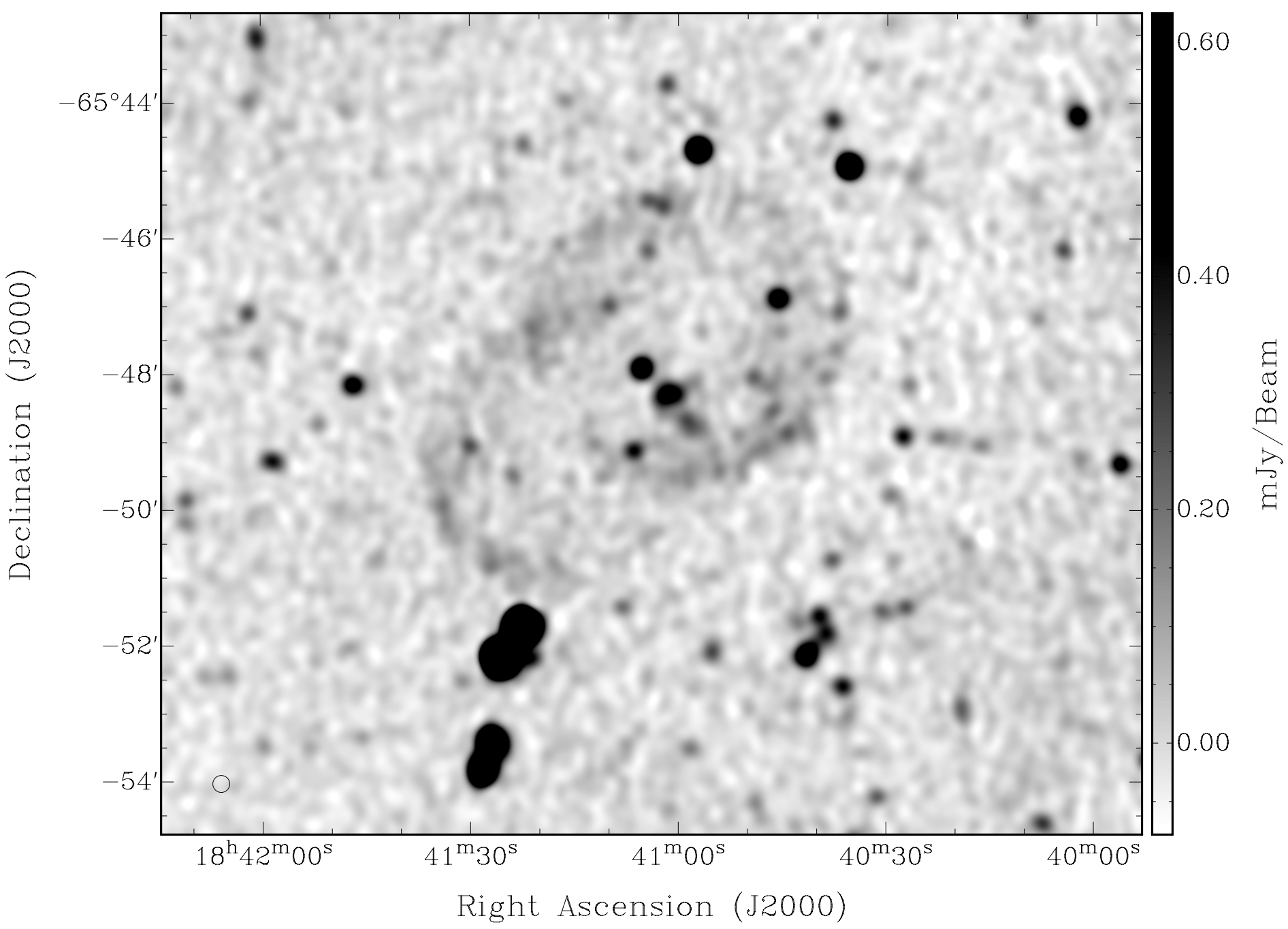}
\caption{Deep (24h) ASKAP 944~MHz radio continuum image of the double-shell system ORC J1841--6547 (also known as ORC~6). The image resolution (15\arcsec) is indicated in the bottom left corner. }
\label{fig:orc6-greyscale}
\end{figure*}

\begin{figure*}[ht] 
\centering
  \includegraphics[width=17cm]{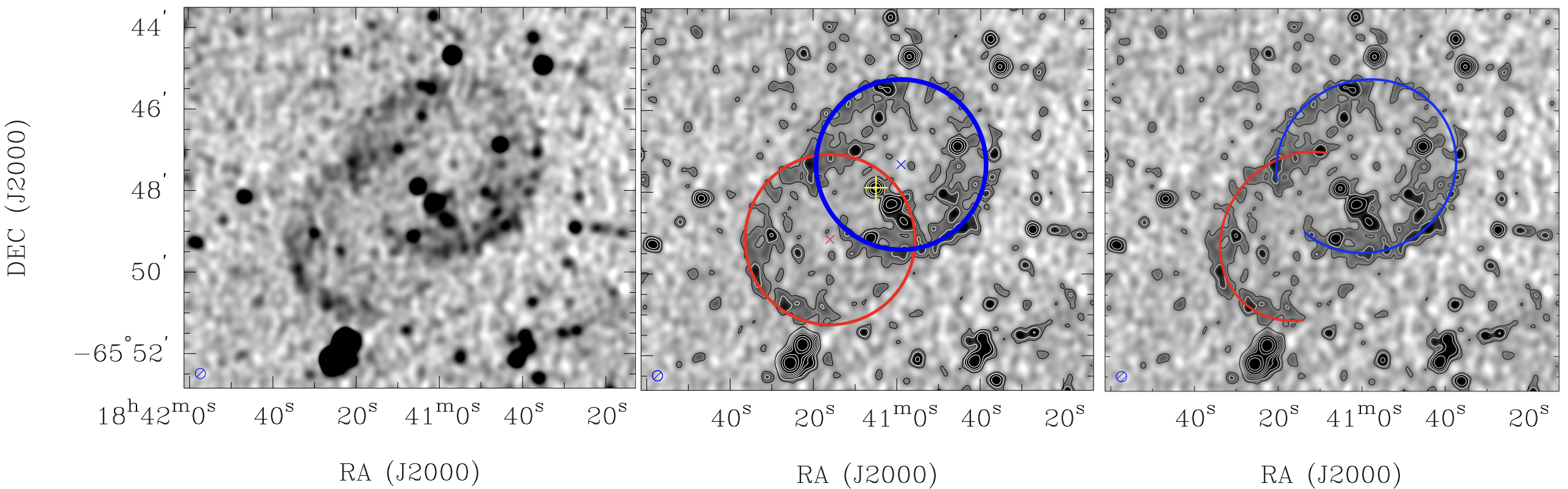}
\caption{The double-shell structure of ORC J1841--6547 as seen in the deep ASKAP 944 MHz radio continuum images. {\bf -- Left:} high-contrast greyscale image. {\bf -- Middle:} overlaid with radio contours at 0.05, 0.012, 0.25, 0.5, 1, 2, 5 and 10 mJy\,beam$^{-1}$ and two colored rings of 240\arcsec\ diameter each. The ring centres are marked with crosses, while the central galaxy is marked with a yellow plus sign. {\bf -- Right:} similar to the middle image, but emphasizing the partial and somewhat elongated nature of the shells. --- The image resolution (15\arcsec) is indicated in the bottom left corner of each panel. }
\label{fig:orc6-collage}
\end{figure*}

In this paper we present the discovery of ORC~J1841--6547 \citep[previously noted as ORC~6 in][]{Dolag2023} found during the search for ORCs and other extended radio sources in a deep $\sim$30~deg$^2$ ASKAP field towards the nearby starburst galaxy NGC~6744 \citep[HIPASS J1909--63a,][]{Koribalski2004}. A summary of the ASKAP multi-epoch observations is given in Section~2, followed by our results in Section~3, comparison with other odd radio circles and radio shell systems in Section~4, and our discussion of the new system in Section~5. Our conclusions are given in Section~6.

\section{ASKAP Observations and Data Processing}
\label{sec:obs}

ASKAP is a powerful radio interferometer consisting of 36 $\times$ 12-m antennas, each equipped with a wide-field Phased Array Feed (PAF) operating at frequencies from 700 MHz to 1.8 GHz \citep{Johnston2008}. Its longest baselines extend to 6.4~km. For radio continuum studies the currently available bandwidth of 288~MHz is divided into $288 \times 1$ MHz coarse channels. A comprehensive system overview is given in \citet{Hotan2021}, and science highlights are presented in \citet{Koribalski2022}.

For the radio analysis of ORC~1841--6547, we obtained three fully calibrated ASKAP 944~MHz radio continuum images from the CSIRO ASKAP Science Data Archive (CASDA)\footnote{CASDA: \url{https://data.csiro.au/domain/casdaObservation}}. 
All fields are centred at $\alpha,\delta$ (J2000) = $19^{\rm h}\,08^{\rm m}\,00^{\rm s}$, --64\degr\,30\arcmin\,00\arcsec, close to the nearby spiral galaxy NGC~6744 \citep[HIPASS J1909--63a,][]{Koribalski2004}. We use scheduling blocks (SB) 32018 (7h), 32039 (7h) and 32235 (10h), observed in Sep 2021, centred at 943.5~MHz; the respective integration times are given in brackets \citep[see also][]{Dobie2023}. The total ASKAP integration time for this field is 24h. ASKAP PAFs were used to form 36 beams arranged in a $6 \times 6$ closepack36 footprint \citep[see][]{Hotan2021}, each delivering a $\sim$30 deg$^2$ field of view out to the half power point. Using rms weighting, we combined the three images after convolving each to a common 15\arcsec\ resolution, achieving a sensitivity of $\sim$25~$\mu$Jy\,beam$^{-1}$ near ORC J1841--6547. Furthermore, we were able to make an ASKAP 1.4~GHz radio continuum image at 20\arcsec\ resolution by combining five 1h images from SBs 43495, 43530, 43570, 43605 and 43695, also available in CASDA. 
Here the centre frequency is 1367.5~MHz, but the bandwidth is only 144~MHz due to interference in part of the band. We measure an rms noise of $\sim$80~$\mu$Jy\,beam$^{-1}$.

\begin{table}
\centering
\caption{Properties of ORC J1841--6547}
\begin{tabular}{ll}
\hline
\hline
  radio morphology & double shell system \\
  radio core position & \\
  ~~~~ $\alpha,\delta$ (J2000) & $18^{\rm h}\,41^{\rm m}\,05^{\rm s}.16$, $-65^{\circ}\,47\arcmin\,54.6\arcsec$ \\
  ~~~~ Galactic $l,b$ & 329.5\degr, --23.5\degr \\
  radio core flux 
    & 1.13 mJy (944~MHz) \\
    & 0.75 mJy (1368~MHz) \\          
  host galaxy name & WISEA J184105.19--654753.8 \\
  host galaxy redshift & $\sim$0.18 
   (see Table~2) \\
  radio shell diameter & $\sim$240\arcsec\ (0.73 Mpc)  \\
  radio system size & $\sim$420\arcsec\ (1.27 Mpc)  \\
  radio shell flux 
     & 14.3 mJy (944~MHz) \\
     & 6.0 mJy (1368~MHz) \\
  environment & compact galaxy group \\
\hline
\end{tabular}
\label{tab:radio-orc}
\end{table}

\begin{figure}[ht] 
\centering
  \includegraphics[width=8cm]{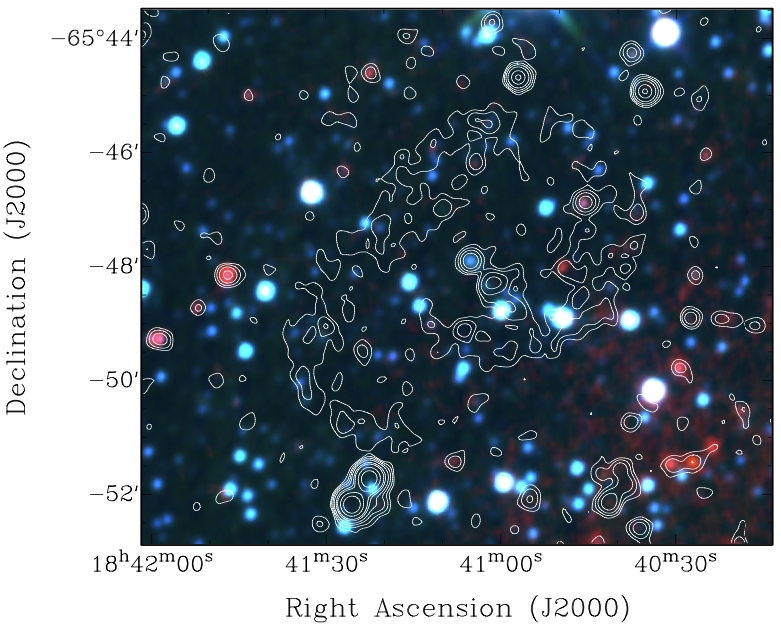}
\caption{ASKAP 944~MHz radio continuum contours of ORC J1841--6547 overlaid onto an RGB colour image created from the WISE infrared bands. The image resolution and radio contour levels are as in Figure~\ref{fig:orc6-collage}.}
\label{fig:orc6-wise-rgb}
\end{figure}

\section{Results}

The double-shell system, named ORC J1841--6547, was found in ASKAP 944~MHz radio continuum images and is shown in Figure~1. It consists of two overlapping, partial shells or bubbles spanning a total extent of $\sim$7\arcmin. The system properties are summarised in Table~1. Despite their low surface brightness, the radio shells are well defined against their surroundings. In Figure~2 we show a high contrast radio image of ORC J1841--6547 as well as coloured overlays to trace its double-shell morphology, followed by a WISE infrared image overlaid with ASKAP contours in Figure~3. More details are given in Section~3.1. 

Within the shell intersect region we note three adjacent radio sources. Of these, the compact, central radio source is associated with the elliptical galaxy WISEA J184105.19--654753.8. This is the likely host galaxy of ORC J1841--6547 with a photometric redshift of $\sim$0.18 \citep{WH2024}, depicted in Figures~4 \& 5. More details are presented in Section~3.2. The two extended radio sources SW of the host are discussed in Section~3.3, followed by a study of the system's potential X-ray emission in Section~3.4.

\subsection{The radio shell emission} 

The stunning double ring radio morphology of ORC~J1841--6547 consists of a near-complete (primary) ring centred $\sim$60\arcsec\ NW of the central radio source, associated with a massive elliptical galaxy discussed in the next section, and a partial (secondary) ring, centred $\sim$90\arcsec\ SE of the host. The NW ring has only a small gap towards the south-east, while only half of the SE ring is currently detected (see Figure~2). The outer ring diameters are well defined (each $\sim$240\arcsec\ in size), seen in clear contrast against the surrounding image noise. Some ring segments appear to be very narrow (not resolved by the 15\arcsec\ ASKAP synthesized beam), while other segments are more diffuse and much wider than the beam. Fitting a polynomial to the NW ring profile gives a mean ring width of 20\arcsec.

Figure~2 shows a high contrast ASKAP radio continuum image of the ORC J1841--6547, overlaid with two rings (middle panel) and two partial, slightly elliptical segments ($PA \approx 45^{\circ}$, right panel) following the brightest radio emission. Figure~3 shows the ASKAP radio contours overlaid onto a WISE colour-composite image. Both rings are likely radio shells or bubbles where the diffuse emission detected inside their outer rims is mainly from the curved shell surface seen in projection. We look into several formation mechanisms in Section~4.

The total ASKAP 944~MHz flux density, $S_{\rm 944MHz}$, of the system is $17.5 \pm 0.5$~mJy. This includes $3.2 \pm 0.2$~mJy from the central region shown in Figure~4, encompassing the radio source associated with the host galaxy and the south-western extension, as well as 0.9~mJy from the radio-bright background galaxy (WISEA J184045.77--654653.2) located within the NW ring. This means we detect $\sim$13.6~mJy in the radio shells alone, which gives a radio power of $P_{\rm 944MHz} = 4~\pi~D_{\rm L}^2~S_{\rm 944MHz}$ = $1.2 \times 10^{24}$ W\,Hz$^{-1}$ for $D_{\rm L}$ = 872~Mpc (for $S_{\nu} \propto \nu^\alpha$). 

The 2\arcmin-resolution GLEAM 200~MHz images \citep{Hurley-Walker2017}, which typically have a 5$\sigma$ detection threshold of 50~mJy\,beam$^{-1}$, do not allow a flux estimate due to confusion. We calculate expected 200~MHz shell fluxes of $\sim$65, $\sim$300~ and $\sim$1500~mJy for $\alpha = -1, -2$ and --3, respectively.



The ASKAP 1.4~GHz images of ORC~6 are very noisy compared to those at 944~MHz, which means that our 1.4~GHz flux estimates are very uncertain. By integrating over the area detected at 944~MHz we measure a 1.4~GHz flux of $\sim$6~mJy corresponding to a radio power of $P_{\rm 1.4GHz} = 5.5 \times 10^{23}$ W\,Hz$^{-1}$. This includes the core, the wedge-shaped emission and the radio shells. The latter have a total flux of $\sim$4.0~mJy ($P_{\rm 1.4GHz} = 3.8 \times 10^{23}$ W\,Hz$^{-1}$), which suggests a steep spectral index of $\alpha \sim -3$. 
We discuss potentially associated X-ray emission in Section~3.4.

\begin{figure}[ht] 
\centering
  \includegraphics[width=8.5cm]{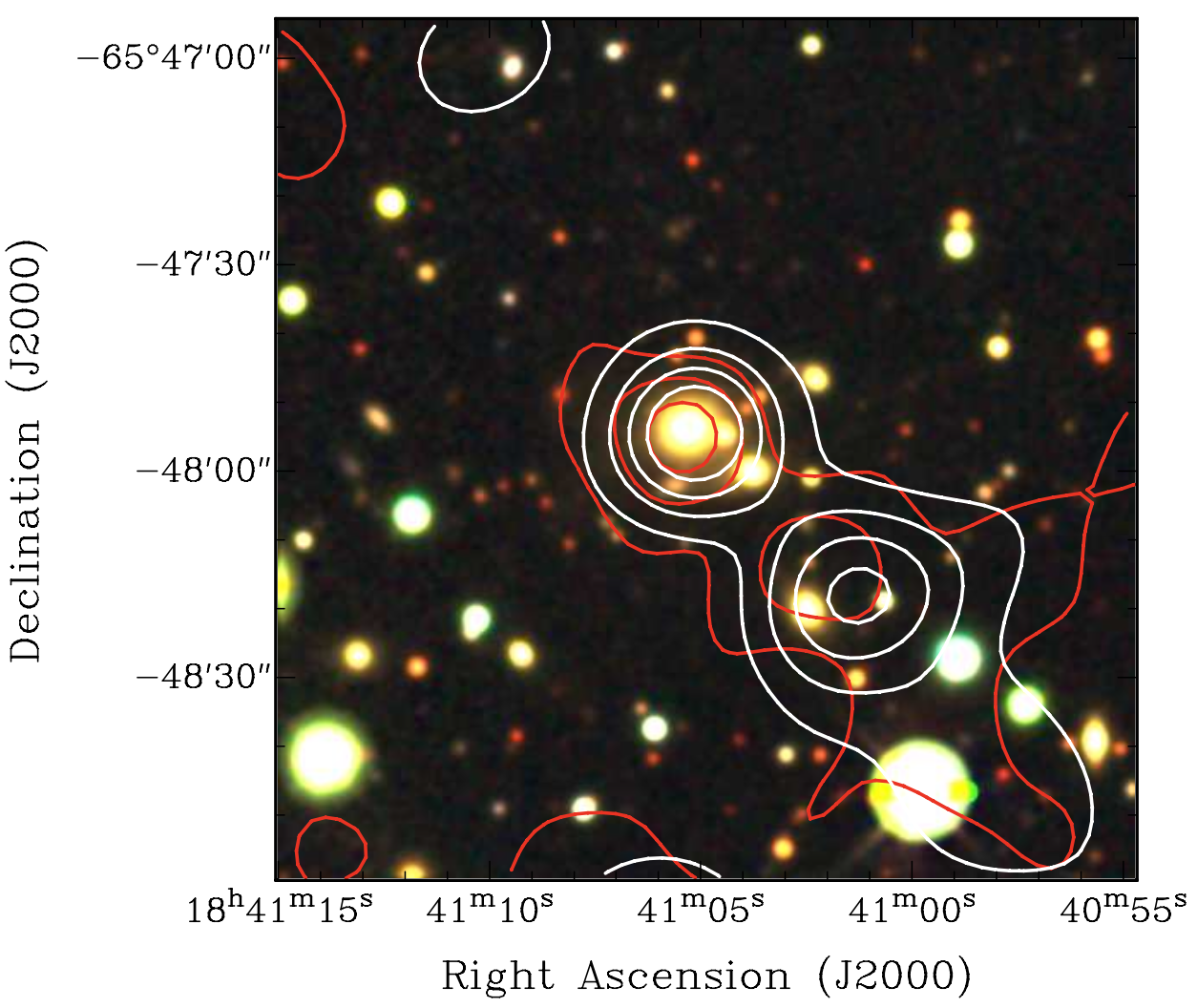}
  \includegraphics[width=8.5cm]{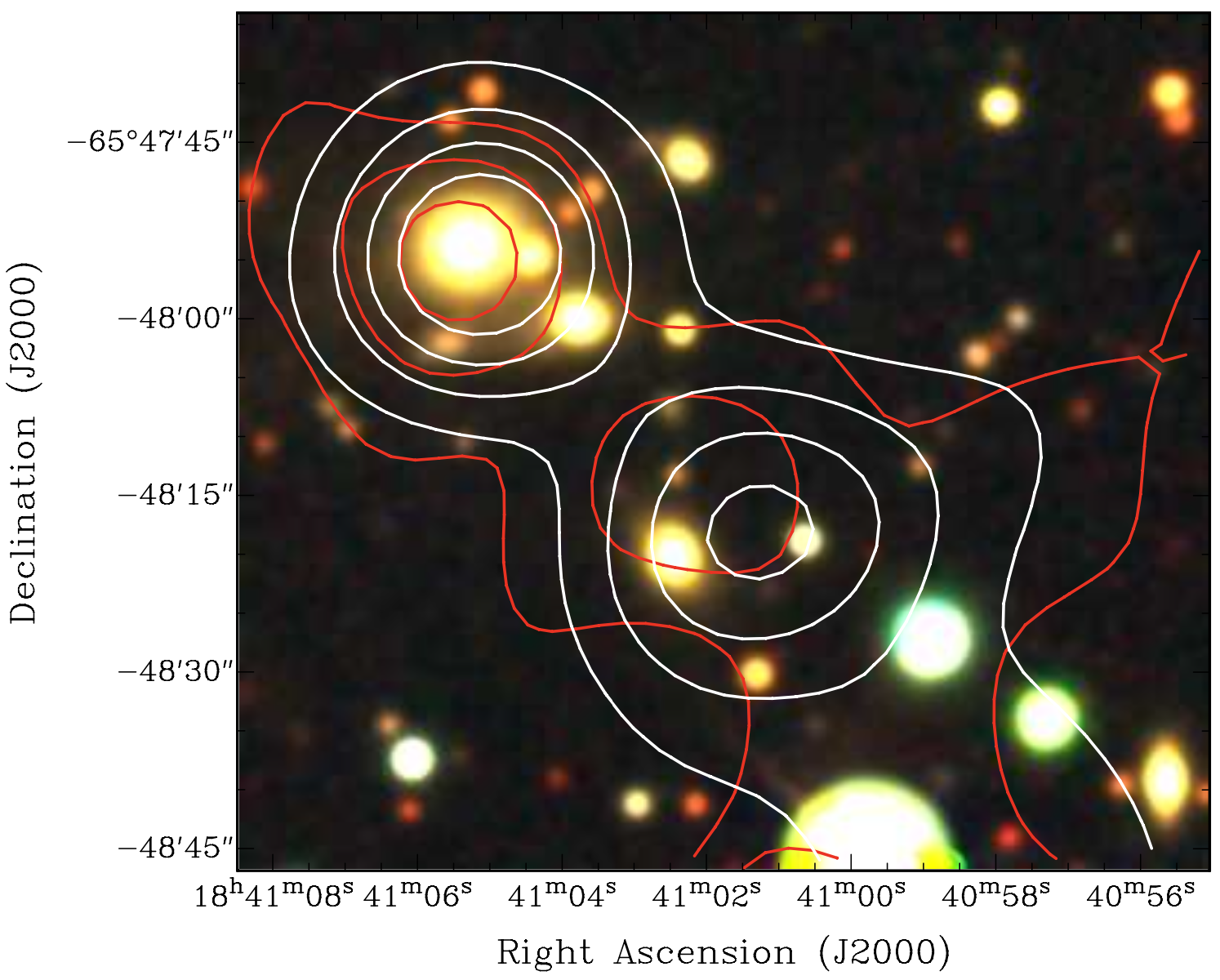}
\caption{The inner region of ORC J1841--6547. --- DESI Legacy Survey DR10 optical RGB ($irg$-bands) image of the ORC J1841--6547 host galaxy and its surroundings overlaid with ASKAP radio contours at 944~MHz (white: 0.2, 0.4, 0.6 and 0.8 mJy\,beam$^{-1}$; 20\arcsec\ resolution) and 1.4~GHz (red: 0.2, 0.4 and 0.6 mJy\,beam$^{-1}$; 20\arcsec\ resolution). The {\bf top} image is centred on the host galaxy, while the {\bf bottom} image zooms in to the area SW of the host galaxy. }
\label{fig:orc6-host}
\end{figure}

\begin{figure}[htb] 
\centering
  \includegraphics[width=8.5cm]{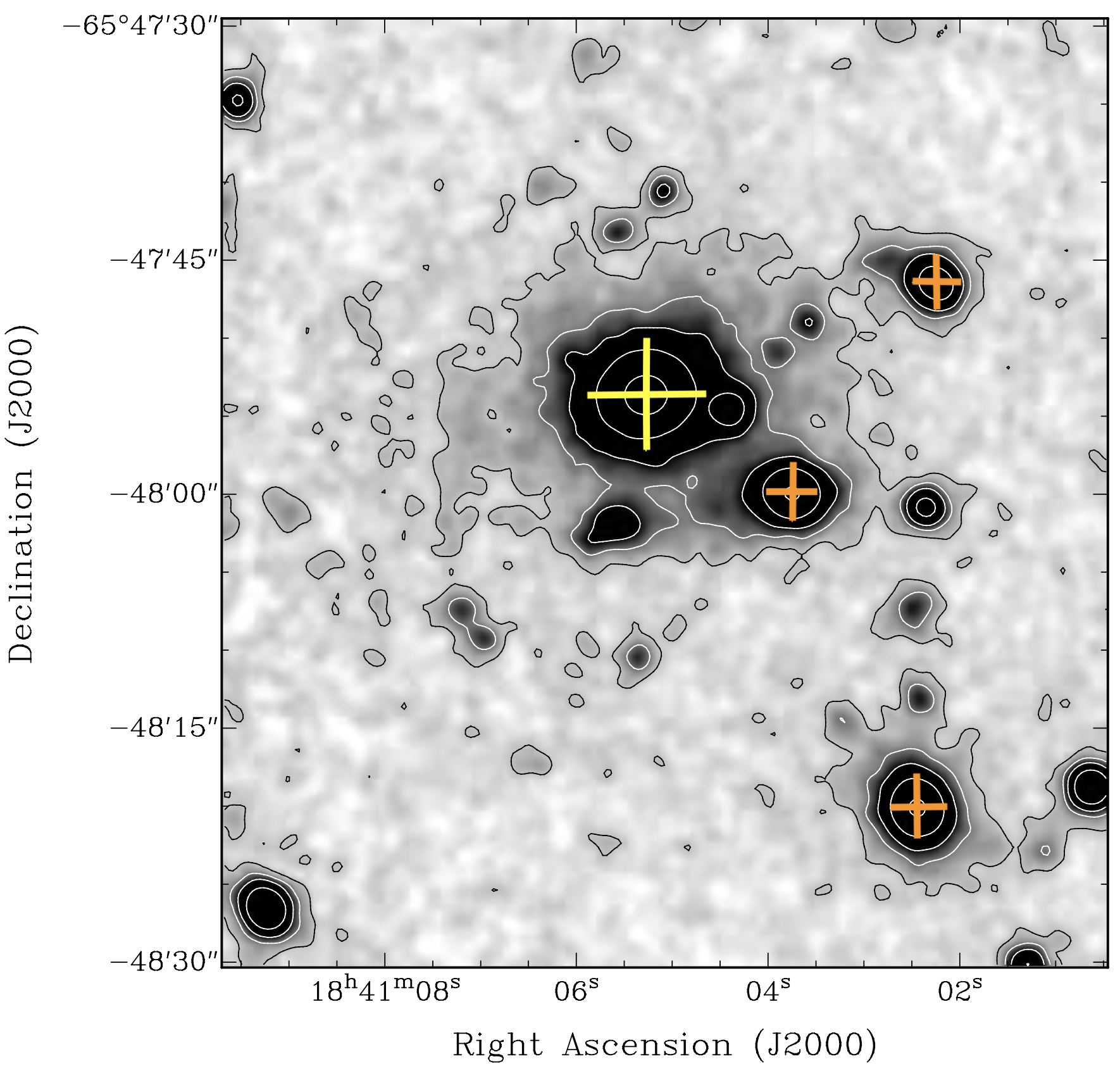}
\caption{DESI Legacy Survey DR10 optical $g$-band image of the ORC J1841--6547 host galaxy (marked with a yellow cross) and surroundings. Three likely companion galaxies, based on their photometric redshifts (see Table~4), are marked with orange crosses. Diffuse stellar light is highlighted with a black contour while white contours show the brighter stellar bodies.}
\label{fig:orc6-group}
\end{figure}

\subsection{The host galaxy and its environment} 

The central, compact radio source in the intersect of the double shell system is associated with the elliptical galaxy WISE J184105.19--654753.8 (see Figure~4). This is the likely host galaxy of ORC J1841--6547. Its radio flux density is 1.13 mJy at 944~MHz and 0.75~mJy at 1.4~GHz, resulting in a spectral index of $\alpha = -1.1$ for $S_{\nu} \propto \nu^\alpha$ (see Table~4).
A summary of the galaxy properties is given in Table~2. Its classification as an elliptical galaxy is based on both the galaxy's optical appearance in the DESI Legacy Survey images \citep{Dey2019} and its location in the WISE colour-colour diagram \citep[see][]{Jarrett2017}. From its 2MASS $K_{\rm s}$-band luminosity we estimate a black hole (BH) mass of $\sim$4 $\times 10^8$\Msun, following Graham (2007). For the host galaxy, WISE J184105.19--654753.8, \citet[][hereafter WH24]{WH2024} estimate a photometric redshift of 0.18, based on DESI Legacy Survey images \citep{Dey2019}, and derive a stellar mass of log $M_{\star}$ = 11.5. Examining the redshift catalog by WH24 we find three likely companion galaxies (listed in Table~3), establishing the ORC host galaxy as the brightest galaxy of a compact group. An envelope of diffuse light, spanning at least 30\arcsec\ (70 -- 90~kpc, see Figure~5) surrounds the host galaxy and its closest companion (c1), indicating tidal interaction and/or merger activity.  WH24 note that WISE J184105.19--654753.8 has the attributes of a brightest cluster galaxy (BCG). They give a cluster membership of 12 galaxies, a radius of $R_{500c}$ = 0.6~Mpc and a mass of $M_{500c} = 0.75 \times 10^{14}$\Msun\ (both for the enclosed overdensity 500 times larger than the critical density of the Universe). The location of ORC J1841--6547 in the radio power vs total mass diagram, shown by \citet[][their Figure~6]{Koribalski2024-Physalis}, is between ORCs~1 and 4, on the low-mass side of \citet{Pasini2022}'s brightest cluster radio galaxy sample (see also Veronica et al. 2025, submitted).

\begin{table} 
\centering
\caption{Properties of the ORC J1841--6547 host galaxy and associated galaxy group}
\begin{tabular}{ll}
\hline
\hline
 DESI DR10 position 
 & $18^{\rm h}\,41^{\rm m}\,05.27^{\rm s}$, 
   --65\degr\,47\arcmin\,53.6\arcsec \\
 & 280.27197\degr, --65.79823\degr \\
  galaxy names 
  & DES J184105.27--654753.64 \\ 
  & 2MASS J18410527--6547536 \\
  & 2MASX J18410522--654753 \\
  & WISEA J184105.19--654753.8 \\
  & SMSS J184105.10--654753.7 \\
  photometric redshifts 
    & 0.1248$^{\rm a}$ \\
    & 0.1463$^{\rm b}$ \\
    & $0.1795 \pm 0.0052^{\rm c}$ \\
    & 0.1784$^{\rm c}$ \\
%
  galaxy stellar mass$^{\rm c}$ & $3.1 \times 10^{11}$ M$_{\odot}$ \\ 
  BCG-like galaxy$^{\rm c}$ & yes \\
  $R_{500c}$ radius$^{\rm c,*}$ & 0.603 Mpc \\
  $M_{500c}$ mass$^{\rm c,*}$ & $0.75 \times 10^{14}$\Msun \\
  2MASS$^{\rm d}$ $K_{\rm s}$ [mag] 
   & $14.45 \pm 0.11$ (profile fit) \\
   & $14.04 \pm 0.1$ ($4''$ aperture) \\
  $->$ black hole mass & $\sim$4 $\times\ 10^8$ M$_{\odot}$ \\
  WISE$^{\rm e}$ W1,2,3,4 [mag]
   & 13.71, 13.55, 12.73, $>$9.26 \\
  DESI DR10 $griz$ [mag] & 17.96, 16.74, 16.29, 15.97 \\
  ASKAP flux density 
   & $1.10 \pm 0.02$ mJy (944~MHz) \\
   & $0.75 \pm 0.11$ mJy (1.4~GHz)\\
\hline
\end{tabular}
{\flushleft References: (a) \citet{Bilicki2014}, (b) \citet{Bilicki2016}, (c) \citet{WH2024}, (d) \citet{Cutri2013}
(e) profile fit, \citet{Wright2010}, fluxes are possibly confused by two neighbouring sources. --- $^{*}$group properties.}
\label{tab:orc-host}
\end{table}

\subsection{The jet-like emission} 

Radio continuum emission is also detected to the SW of the host galaxy (see Figure~4), extending over $\sim$70\arcsec. While two of the three likely companion galaxies (see Figure~5) are located within this emission patch, the radio peaks are offset from any galaxies. In particular, the brightest radio component (944~MHz peak at 18:41:01.3, --65:48:17, $\sim$30\arcsec\ offset from the host galaxy) shows a wedge-shaped morphology resembling an inward moving shock. We note that the 1.4~GHz and 944~MHz peak positions are slightly offset; the respective peak fluxes are $\sim$0.7 mJy\,beam$^{-1}$ and 0.9 mJy\,beam$^{-1}$. Deeper, high-resolution radio continuum images are needed to further analyse this extended structure, which may be the remnant of a jet emerging from the host galaxy. 

\begin{table*} 
\centering
\begin{tabular}{cccccccc}
\hline
 ORC J1841--6547 & $\alpha,\delta$ (J2000) position & DESI DR10 $g, r, i, z$-band & WISE W1, W2 & redshift \zph\ & log stellar mass & notes \\
  & [degr, degr] & [mag] & [mag] & & [\Msun] \\
\hline
\hline
 host galaxy & 280.27197, --65.79823 & 17.728, 16.583, 16.175, 15.883 & 15.918 16.321 & $0.1795 \pm 0.0052$ & 11.492 & BCG-like \\
  c1    & 280.26562, --65.79999 & 19.361, 18.212, 17.808, 17.497 & 17.618, 18.050 & $0.1742 \pm 0.0051$ & 10.662 \\
  c2    & 280.26019, --65.80558 & 19.234, 18.046, 17.618, 17.297 & 17.315, 17.721 & $0.1773 \pm 0.0045$ & 10.822 \\
  c3    & 280.25940, --65.79623 & 20.464, 19.357, 18.983, 18.704 & 19.024 19.899 & $0.1884 \pm 0.0069$ & 10.065 \\
\hline
\end{tabular}
\caption{Properties of the ORC J1841--6547 host galaxy and its brightest companion galaxies (c1, c2, and c3), based on their photometric redshifts by \citet{WH2024}, obtained from the DESI Legacy Imaging surveys DR10.}
\label{tab:group-prop}
\end{table*}

\begin{table}
\centering
\begin{tabular}{cccccc}
\hline
& 1.4 GHz flux & 944 MHz flux & $\alpha$  \\
& [mJy] & [mJy]  \\
\hline
\hline
host galaxy & $0.75 \pm 0.11$ & $1.13 \pm 0.08$ & --1.0 \\
host + ext. & $1.9 \pm 0.3$ & $3.2 \pm 0.2$ & --1.3 \\
NW shell (blue) & $2.9 \pm 0.4$ & $8.7 \pm 0.2$ & --2.8 \\ 
SE shell (red) & $1.3 \pm 0.4$ & $4.9 \pm 0.2$ & --3.4 \\
total & $\sim$6 & $17.5 \pm 0.5$ & --2.7 \\
\hline
\end{tabular}
\caption{ASKAP radio continuum flux densities for the components of ORC~J1841--6547. }
\label{tab:radio-flux}
\end{table}



\subsection{Diffuse X-ray emission} 

We find a catalogued X-ray source, possibly associated with ORC~J1841--6547, in both ROSAT \citep[1WGA J1840.8--6546,][]{WGA2000} at $\alpha,\delta$ (J2000) = $18^{\rm h}\,40^{\rm m}\,50.3^{\rm s}$, --65\degr\,46\arcmin\,38\arcsec\ with a 0.2--2 keV energy of $1.53 \times 10^{-13}$ erg\,s$^{-1}$\,cm$^{-1}$
and eROSITA \citep[1eRASS J184048.9--654700,][]{Merloni2024} with $7.2 \times 10^{-14}$ erg\,s$^{-1}$\,cm$^{-1}$ in the 0.2--2.3~keV band. At the adopted redshift of $z = 0.18$ (or luminosity distance $D = 872$~Mpc) this corresponds to an X-ray luminosity of $L_{\rm X}$ = $1.4 \times 10^{43}$ erg\,s$^{-1}$ (ROSAT) and $6.6 \times 10^{42}$ erg\,s$^{-1}$ (eROSITA). 
The position uncertainty for the ROSAT detection is $\sim$1\arcmin\ and for the 1eRASS position $\sim$10\arcsec, and the low count number statistics of the eRASS1 detection preclude robust judgement regarding its spatial extent. The target of the ROSAT pointed observations was the Seyfert galaxy ESO\,103-G035 ($D$ = 58~Mpc), at a projected distance of 28\arcmin\ from the double-shell system.

\begin{figure}[ht] 
\centering
    \includegraphics[width=8cm]{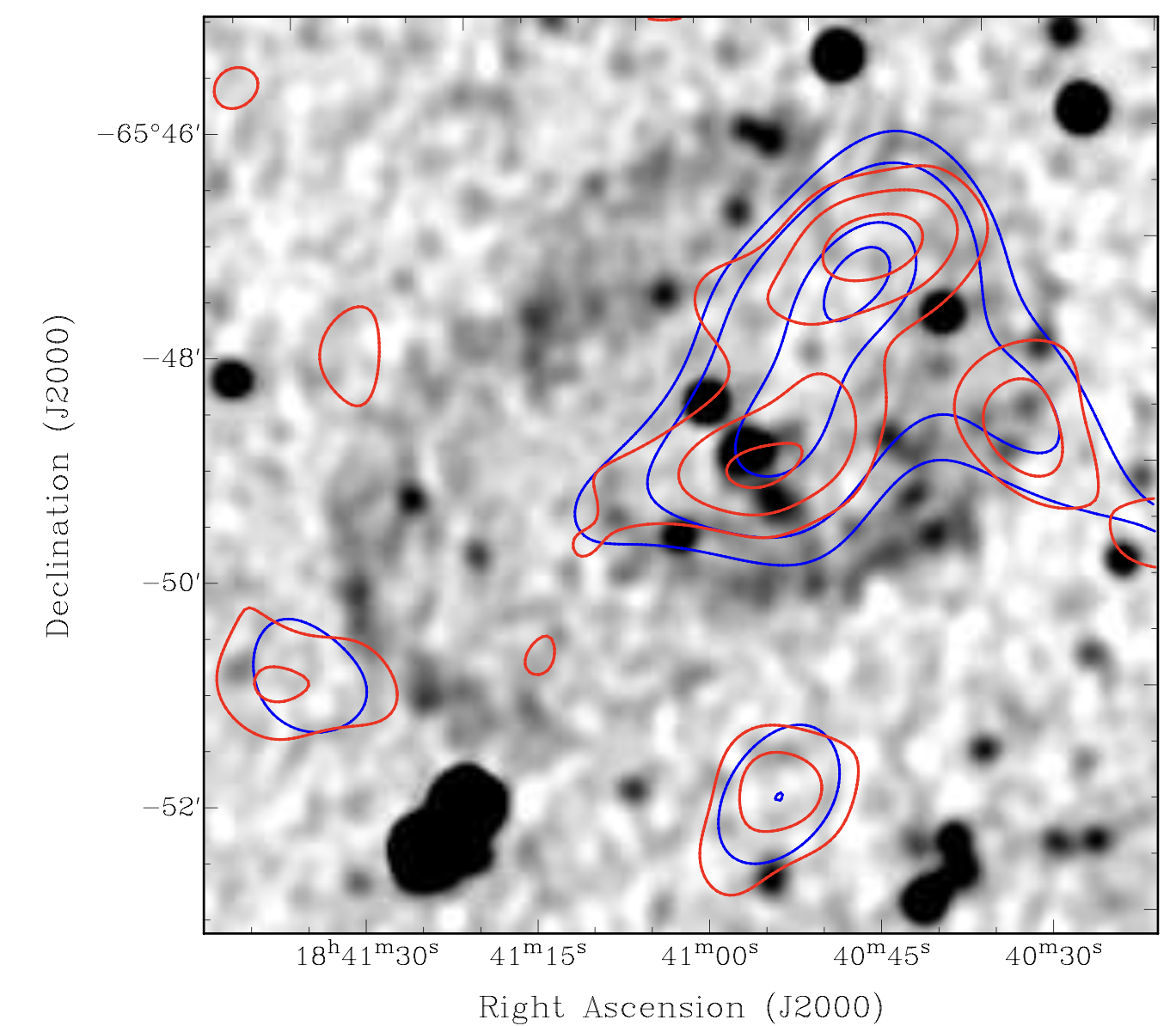}
\caption{ROSAT PSPC X-ray contours overlaid onto the ASKAP 944~MHz radio continuum image (greyscale) of ORC~ J1841--6547. The ROSAT images were smoothed to 60\arcsec\ resolution (red contours) and 100\arcsec\ resolution (blue contours), showing extended X-ray emission within the ORC's north-western shell and overlapping with the host galaxy and radio extension. }
\label{fig:orc6-rosat}
\end{figure}

Figure~\ref{fig:orc6-rosat} shows the ROSAT PSPC X-ray emission contours overlaid onto our ASKAP 944~MHz radio continuum image. The X-ray emission appears to be elongated north-west to south-east, likely consisting of diffuse emission and compact sources. Interestingly, it coincides and partially fills the ORC's north-western shell, while also overlapping with the host galaxy and the south-western radio extension. The position of the south-eastern X-ray maximum is $\sim$1\arcmin\ offset from the host galaxy, coinciding with the radio south-eastern radio extension, while the north-eastern X-ray maximum has no obvious radio or optical counterparts. Deeper X-ray data are needed to confirm and further explore the morphology, extent and possible association of the hot gas with the ORC~J1841--6547 radio shell system and its host galaxy. 

Diffuse X-ray emission, when detected in galaxy groups, is nearly always centred on a luminous elliptical or lenticular galaxy \citep{Mulchaey2003}. Offset X-ray emission likely indicates group members in the process of merging as suggested for the nearby Physalis radio shell system \citep[][see Section~4.2]{Koribalski2024-Physalis}. 

\section{Comparison with other ORCs and radio shell systems}

In the following we briefly compare ORC~J1841--6547 to some of the known single ORCs and two nearby radio shell systems.

\subsection{Comparison to ORCs 1, 4 \& 5}

\begin{figure*}[htb] 
\centering
    \includegraphics[width=17cm]{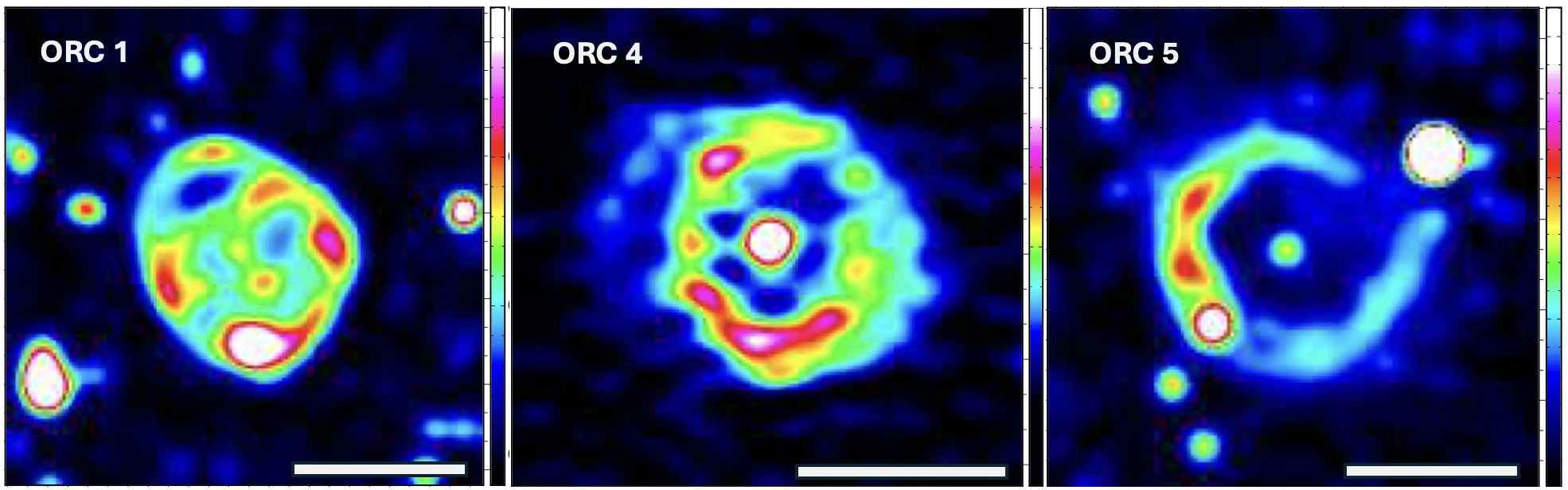}
\caption{MeerKAT 1.3~GHz wide-band radio continuum images of ORC~1 \citep[left,][]{Norris2022}, ORC~4 \citep[middle,][]{Riseley2024}, and ORC~5 (right, Koribalski et al., in prep.). While the ORCs have similar morphologies, the strength the radio core varies substantially as well as the amount of diffuse emission between ring and core. The images are shown at 10\arcsec\ resolution; the length of scale bar is 60\arcsec.  }
\label{fig:meerkat-orcs145}
\end{figure*}

MeerKAT 1.3~GHz radio continuum images of ORCs~1, 4 and 5 are shown in Figure~\ref{fig:meerkat-orcs145}, smoothed to 10\arcsec\ resolution, highlighting the similarities (size, edge-brightening, surface brightness, ...) as well as differences (core brightness, internal structure, ...) in their radio morphologies. Not included in this collage is the recently discovered MIGHTEE ORC \citep{Norris2025}, whose angular size of 35\arcsec\ is about a factor two smaller than that of ORCs~1, 4 and 5 \citep{Norris2021,Koribalski2021}. All four have massive elliptical galaxies at their centres with redshifts of 0.55 (ORC~1), 0.45 (ORC~4), 0.27 (ORC~5) and 0.20 (MIGHTEE ORC), respectively. Deep optical images from the DESI Legacy Imaging Surveys \citep{Dey2019} reveal several, less massive companion galaxies in their surroundings (based on their photometric redshifts), suggesting the central elliptical galaxy is a brightest group galaxy (BGG). For details on the respective ORC group environments, including companions, see the review by Koribalski \& Norris (2025, in prep.). Not further discussed here is ORC~J1027--4422 \citep{Koribalski-Veronica2024}, whose location at low Galactic latitude prevents a detailed study of the likely host galaxy ($z \sim 0.3$).

The radio morphologies of the above ORCs all show edge-brightened radio rings, while ORC~J1841--6547 consists of two partial rings. That said, the high-resolution MeerKAT 1.3~GHz image of ORC~1 by \citet{Norris2022} reveals complex internal structure, likely consisting of two or more rings plus diffuse emission. Furthermore, the GMRT image of ORC~4 reveals a radio segment east of the primary ring, which is now confirmed as a second, much fainter ring in the MeerKAT 1.3~GHz image shown by \citet{Riseley2024} and here in Figure~\ref{fig:meerkat-orcs145}. It is possible that we see a nested ring structure, but its morphology is different from the multi-ring appearance of ORC~1. Projection effects need to be kept in mind, allowing different appearances of similar sources from a changed viewing angle. No secondary ring is seen in 1.3~GHz MeerKAT image of ORC~5, which reveals a pronounced C-shape morphology. We note faint radio synchrotron threads just inside the western ring gap. The SE companion to the host galaxy is very prominently located in the midpoint of the C-shape, giving the structure a WAT-like appearance.

\subsection{Nearby radio shell systems}
Recently, two nearby radio shell systems were found in ASKAP radio continuum images, somewhat resembling the more distant ORCs. This allowed for detailed radio and follow-up X-ray studies. Both systems are characterised by multiple and/or nested radio shells surrounding a central elliptical BGG.  The closest one is the Physalis system \citep[ASKAP J1914--5433,][]{Koribalski2024-Physalis} centred on ESO\,184-G042 and its companion LEDA~418116, both elliptical galaxies, at a distance of only 75~Mpc (redshift $z = 0.017$). Physalis has a diameter of $\sim$145~kpc and features a central radio ridge. Interestingly, our follow-up study with XMM-Newton reveals the X-ray emission to be offset from the radio ridge and centred on the companion galaxy. The second radio shell system is known as the Cloverleaf \citep[ASKAP J1137--0050,][Koribalski et al., in prep.]{Dolag2023} centred on the elliptical galaxy CGCG\,012-043 at a distance of $\sim$200~Mpc (redshift $z = 0.046$), spanning 400~kpc. A follow-up XMM-Newton study by \citet{Bulbul2024} also shows the X-ray emission to be offset. In both cases, the hosts are massive elliptical galaxies surrounded by several companions, i.e. they are BGGs. \citet{Koribalski2024-Physalis} find an example of such offset in Magneticum simulations and discuss possible formation mechanisms.

\section{Discussion}

We established that the host of ORC~J1841--6547 is a massive elliptical galaxy which is surrounded by an envelope of diffuse stellar light, likely the signature of past merger activity, and at least three, much smaller companion galaxies. The host galaxy properties and environment (see Tables~2 \& 3) are similar to those of other ORC host galaxies. Compact radio emission from the ORC host galaxies is detected, but varies in strength (see Figure~\ref{fig:meerkat-orcs145}). While no double-sided radio jets or lobes are observed, past activity of the host's supermassive black hole (SMBH) is likely, suggesting that the galaxy surroundings are filled with the fading (ageing) magnetised plasma of remnant lobes. The low radio surface brightness of ORC~J1841--6547 and absence of active radio jets from the host galaxy suggests the shells are old, most likely relics or remnants left behind after the jet emission ceased \citep[e.g.,][]{KG1994,Shabala2024}. Their edge-brightened morphology strongly suggests that shocks are involved. \\

Overall, a number of physical processes are contributing to the ionized plasma in the CGM of massive elliptical galaxies. The CGM is a highly dynamic, time-variable environment, but not as turbulent as the intra-cluster medium. The apparent lack of radio jets from the central AGN of the ORC J1841--6547 host galaxy suggests it is currently inactive, but former recurring activity means that fossil radio-emitting plasma is present. Depending on the time since last activity, jet driven cavities may be present as well as fading radio lobes whose size, structure and appearance change with age. Stellar and starburst winds as well as turbulence may also be present, traceable via optical spectroscopy \citep[e.g.,][]{Coil2024}. 

Furthermore, the evolution and growth of the massive host galaxy involves accretion and mergers, which on rare occasions drive powerful merger shocks into the CGM \citep{Dolag2023}. These shocks can have a wide range of Mach numbers. Once re-ignited by shock waves, the compressed fossil plasma may again become detectable at radio wavelengths \citep[e.g.,][]{EnsslinBrueggen2002, Riseley2025}. Quantifying the expected radio luminosity is difficult as shown by \citet{Boess2024} due to large uncertainties in CR acceleration efficiency models and magnetic field strength.


\citet{Riseley2025} present a detailed multi-wavelength study of the Hickson Compact Group (HCG) 15, in which no active jets or hotspots are detected from any of the five massive member galaxies. Instead, around galaxies HCG15-C (UGC~1620) and HCG15-D (UGC~1618) extended diffuse radio emission is present with radio powers of $P_{\rm 150MHz} = (3.23 \pm 0.09) \times 10^{23}$ W\,Hz$^{-1}$ and $P_{\rm 1400MHz} = (2.22 \pm 0.02) \times 10^{22}$ W\,Hz$^{-1}$, consistent with that of steep-spectrum remnant radio sources \citep[e.g.,][]{Riseley2022}. Compared to galaxy clusters \citep[e.g.,][]{Venturi2022}, galaxy group environments are characterised by smaller number of galaxies, lower velocity dispersion, lower kinetic energy and less mergers. Merging clusters often show radio relics in their outskirts, created by cluster merger shocks \citep[e.g.][]{Brown2011,vanWeeren2019}. \\

ASKAP has shown to be highly suited to detecting low-surface brightness radio emission, e.g., the ageing lobes of giant radio galaxies \citep{Koribalski2025} and steep-spectrum relic / remnants of unknown nature (Smeaton et al. 2025, sub.). In the following we explore the ORC environments (Section~5.1), discuss the radio power -- mass relation (Section~5.2), re-energised remnant radio lobes (Section~5.3), precessing jets (Section~5.4), other double rings (Section~5.5) and consider ORC formation mechanisms (Section~5.6).



\subsection{Galaxy group environment}

The ORC host galaxies are massive elliptical galaxies surrounded by several companions, suggesting they are BGGs. While they resemble BCGs, their masses and number of companions are lower than for clusters. Galaxy groups typically host several large galaxies, which dominate the overall mass and luminosity, and many dwarf galaxies, which can be excellent tracers of the group dynamics, interactions and total mass. Groups dominated by spiral galaxies are typically gas-rich, see for example H\,{\sc i} studies of the Hickson Compact Group (HCG) 44 \citep{Serra2013} and the NGC~6221 group \citep{KD2004}, while those dominated by elliptical galaxies, contain little gas. \citet{OP2004} find that most X-ray bright groups contain a bright central early-type galaxy and highlight the importance of elliptical BGGs. In a follow-up study of 30 nearby groups in their sample, \citet{Croston2005} find 19 (63\%) to be associated with a AGN-related radio source, with roughly half of these in an active state (i.e., showing a double-lobed structure). Most of the currently inactive BGGs would have gone through periods of SMBH activity; their remnant lobes now either dispersed or only detectable in deep radio surveys. Although, \citet{Giacintucci2011} find that the intragroup medium (IGM) can confine such remnant lobes, preventing them from dissipating quickly.

An example of a remnant/relic radio galaxy is NGC~1534, studied in detail by \citet{Hurley-Walker2015} and \citet{DuchesneHollitt2019}. It shows two fading lobes, spanning 600~kpc, with a very steep spectral index ($\alpha\ = -2.1 \pm 0.1$). The radio lobes are very diffuse with no signs of jets or hot spots. The host, NGC~1534, is a lenticular galaxy with a prominent dust lane (like NGC~5128) at a distance of 80~Mpc (\zsp\ = 0.017816). It is the BGG of a loose group. Another example is the nearby NGC~507 galaxy group \citep{Brienza2022}, which shows the presence of diffuse radio emission with complex, filamentary morphology likely related to a previous outburst of the central galaxy. 
Here the remnant plasma has been displaced by the sloshing motions on large scales, likely caused by interactions between remnant plasma and the external medium. 

\begin{figure*}[h] 
\centering
    \includegraphics[width=12cm]{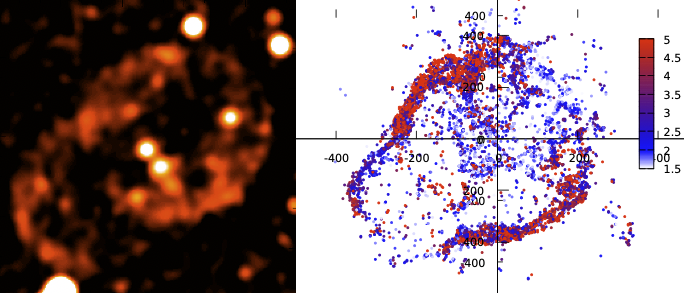}
    \includegraphics[width=6cm]{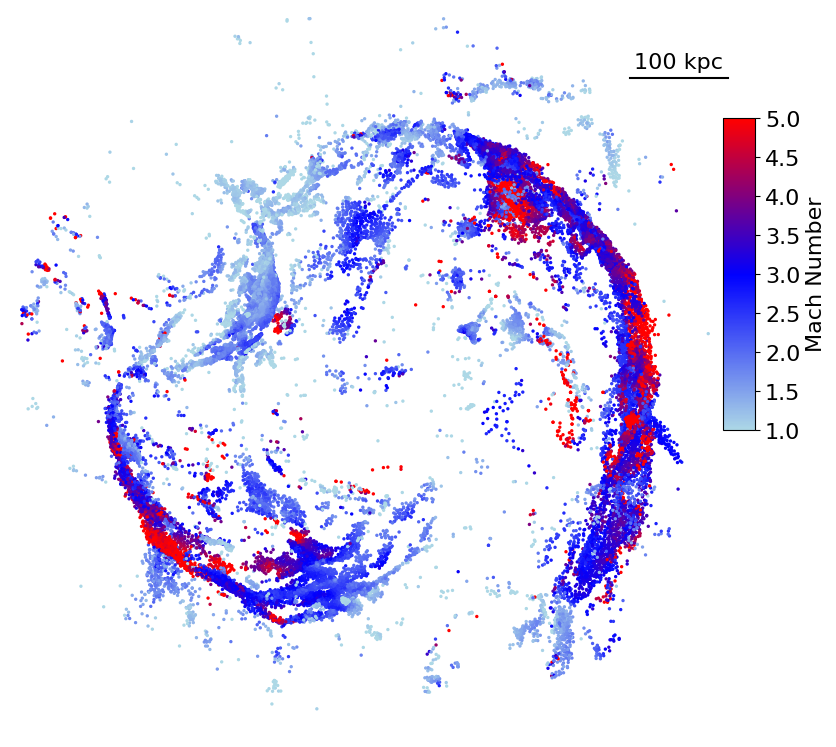}
\caption{{\bf -- Left:} ASKAP 944~MHz radio continuum image of ORC~J1841--6547 (ORC~6). {\bf -- Middle and right:} Simulated merger shocks from a MW galaxy like halo \citep[middle,][]{Dolag2023} and a ten times more massive halo \citep[right,][]{Ivleva2025}, both selected to roughly match the morphology of ORC~6. The colour coding on the middle and right panels show the sonic Mach number determined by the shock finding algorithm \citep{Boess2024}.}
\label{fig:orc6+sims}
\end{figure*}

In addition, the CGM filling the space between the currently inactive BGGs, which is mixed with the ageing electron plasma from remnant radio lobes, is also subject of dynamical processes driven by the still ongoing assembly of the groups. Simulations demonstrate that powerful merger shocks, e.g. formed on rare occasions during the evolution of the BGG, will expand through the enriched CGM. This happens in an analogous way to that in galaxy clusters, but due to the lower density and temperature of the CGM this cannot be directly observed in X-rays as the counterparts in galaxy clusters. Thereby these shock waves are sweeping up and re-energising cosmic rays in the diffuse plasma, resulting in edge-brightened shells detected as radio rings, aka ORCs. A wide range of ORC sizes and morphologies can result in haloes of varying sizes. Figure \ref{fig:orc6+sims} shows some examples for this scenario, compared to ORC~6. The middle panel shows an example for a Milky Way-sized halo, which underwent a triple merger event \citep[see][]{Dolag2023}. The right panel shows a ten times more massive halo, undergoing a major merger event \citep[see][]{Ivleva2025}. In such scenarios, various complex morphologies can appear, including double shells and nested shells. Interestingly, the shallower potential in a group-like environment together with the impact of AGN feedback can lead to more significant offsets between the X-ray bright CGM and the BCG \citep{Koribalski2024-Physalis, Ivleva2025}. Here, the X-ray bright CGM is sometimes centered on the impacted galaxy instead of the BGG. The Physalis ORC is the only one that is close enough to have very significant X-ray observations showing this offset \citep{Koribalski2024-Physalis}, while for Cloverleaf and ORC~6 they only have indications of an offset diffuse emission in X-rays.


\begin{figure*}[h] 
\centering    
   \includegraphics[height=5.1cm]{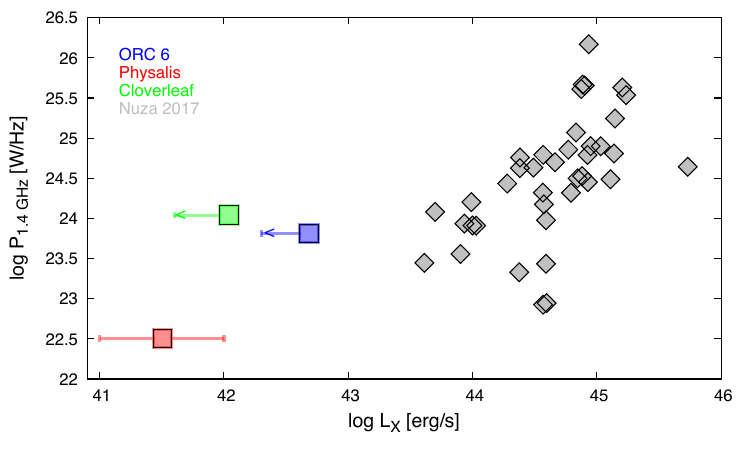}
   \includegraphics[height=5.2cm]{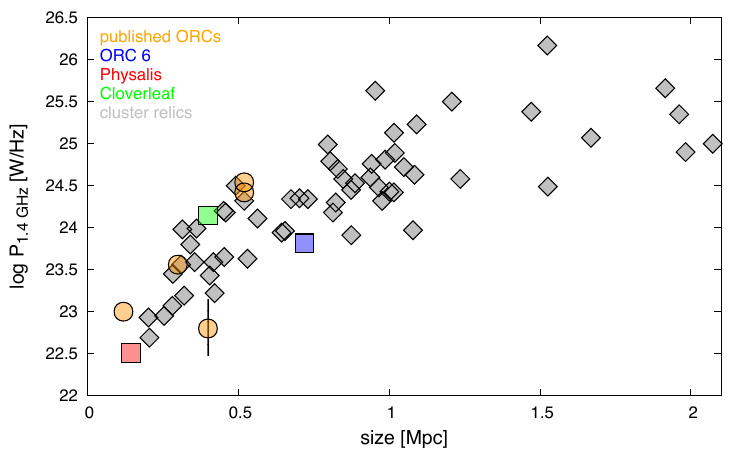}
\caption{Scaling relation of 1.4~GHz radio power ($P_{\rm 1.4GHz}$) vs X-ray luminosity (left) and size (right) for 39 NVSS-detected clusters with radio relics from \citep{Nuza2017} and for ASKAP-detected radio shell systems in three nearby galaxy groups. For the radio relics we use their catalogued NVSS 1.4~GHz flux densities and largest linear sizes (LLS). For the radio shell systems, Cloverleaf (green; $D = 200$~Mpc), Physalis (red; $D = 75$~Mpc) and ORC~6 (blue, $D = 872$~Mpc), we use the measured 1.4~GHz flux densities of the radio shells and the respective ring/shell diameters. On the right side we added five known ORCs (orange) for which we have no X-ray detections. From smallest to largest these are: the MIGHTEE ORC, ORC~5, ORC J1027--4422 (uncertain redshift), plus ORCs 1 and 4. We use the respective MeerKAT 1.3~GHz images to measure their approximate ring fluxes.} 
\label{fig:LXP1400-relation}
\end{figure*}


\subsection{Scaling relations}

The location of ORC J1841--6547 (ORC~6) in the radio power vs total mass diagram, shown by \citet[][their Figure~6]{Koribalski2024-Physalis}, is between ORCs~1 and 4, on the low-mass side of \citet{Pasini2022}’s brightest cluster radio galaxy sample.

In Figure~\ref{fig:LXP1400-relation} we explore the scaling relation between the 1.4~GHz relic radio power and the X-ray luminosity for galaxy clusters and compare them to the three radio shell systems where X-ray data are available. We took the set of 39 galaxy clusters, which have been uniformly extracted from the NRAO VLA Sky Survey (NVSS) by \citet{Nuza2017}. Some of the systems have several identified features (like double relics), which we then summed together to have a fair comparison to the total radio luminosity of the shells within the ORC systems. While the Physalis system seems to be in an early stage of the evolution within these systems, ORC~6 and the Cloverleaf system have comparable size and radio power than the relics in galaxy clusters, despite being much less massive. The similarity in size indicates that the shallower potential within the group environment, allows the powerful shocks to travel to relatively larger distances which in turn can make them more circular. In addition, reaching the same power despite being much less massive indicates that the relative contribution of the AGN activity foster much more effective conversion of the shock energy into radio emission, similar to the relatively larger impact of AGN feedback within groups compared to galaxy clusters.



A detailed radio and X-ray study of the Physalis system was carried out by \citet{Koribalski2024-Physalis}, who measure total ASKAP 944~MHz and 1.4~GHz flux densities of at least $145 \pm 2$~mJy and $79 \pm 2$~mJy, respectively. These correspond to radio powers of $P_{\rm 944MHz} = 9.8 \times 10^{22}$~W\,Hz$^{-1}$ and $P_{\rm 1.4GHz} = 5.4 \times 10^{22}$~W\,Hz$^{-1}$. The Physalis radio shells were estimated to contain $\sim$60\% of the total flux, corresponding to $3.2 \times 10^{22}$~W\,Hz$^{-1}$ used in Figure~9. An analysis of new XMM X-ray data is under way (Khabibullin et al., in prep.).

\subsection{Remnant radio lobes}

The radio morphology of ORC J1841--6547 somewhat resembles that of nearly circular (fat) double-lobed radio galaxies such as Fornax~A \citep{Fomalont1989} but only the radio cocoon, created by the plasma backflow, is detected. Neither active inner jets nor hot spots are seen. While such remnant radio galaxies are quite common, their lobes are rarely edge-brightened or devoid of internal emission. But there is evidence for limb-brightening at the radio lobe boundaries \citep[e.g.,][]{Carvalho2005, Daly2010} such that FR\,II lobes may indeed be radio hollow \citep{MathewsGuo2012}.

Simulations of the remnant lobe evolution of FR\,II-type radio galaxies by \citet{Shabala2024} show significant edge-brightening when re-energised by a plane-perpendicular shock wave. Alternately, merger shocks propagating outwards from the central galaxy (at rare occasions during its evolution) into remnant lobes will also re-ignite the magnetised plasma. There may be consecutive merger shocks contributing to the radio brightness of the shells or bubbles during the growth of the central elliptical.

On smaller scales, buoyant bubbles in 2D and 3D simulations \citep[][respectively]{Churazov2001,ONeill2009} expanding into the ICM can form vortex rings or torus-like features when viewed at or close to the jet axis \citep[see also][]{Brienza2021}.

\subsection{Precessing radio jets}
Could the two rings of ORC~J1841--6547 have been created by double-sided precessing jets emerging from the central galaxy\,?
Geometrically, the time-integrated 3D structure of such a system would resemble a rim-brightened double cone with a wide opening angle, resembling an hour-glass slightly inclined to the line of sight. The latter looks like a single ring if viewed face-on. The peculiar radio extension SW of the host galaxy, shown in Figure~\ref{fig:orc6-host}, could be related to an active one-sided jet. If the two shells were from an episode of precessing jet activity, their brightness is surprisingly uniform. None of the simulated shape seen by \citet{Horton2020} resemble ORC~J1841--6547.

In a recent paper, \citet{Nolting2023} present simulations of jet precession in radio galaxies to examine how they evolve over time and model their radio synchrotron emission. They find that the jet trajectories can become unstable due to their own self-interactions and lead to "reorientation events", often observed as X-, S- and Z-shaped radio galaxies. Another cause of the change in the jet direction might be associated with the spin evolution of the accreting BH, and cosmological simulations tracking the spin evolution in a self-consistent manner show that rapid spin direction changes are not uncommon \citet{Sala2024}.

The synthetic radio intensity and spectral index maps (and movies) produced by \citet{Nolting2023} at different viewing angles can be used for comparison with high-resolution radio observations. ORC-like structures are seen when viewing the remnant radio emission of precessing jets approximately end-on, shortly after the jet power has been turned off. The low brightness, specific timing and viewing angle would make these very rare radio morphologies in agreement with observations. \citealt{Nolting2023}'s 3D magneto-hydrodynamical simulations allow for both double and single ring morphologies. Polarisation maps would be of particular interest and should allow to distinguish this scenario from radio relics created by merger shocks \citep{Dolag2023}. 


\begin{figure}
\centering
    \includegraphics[width=8cm]{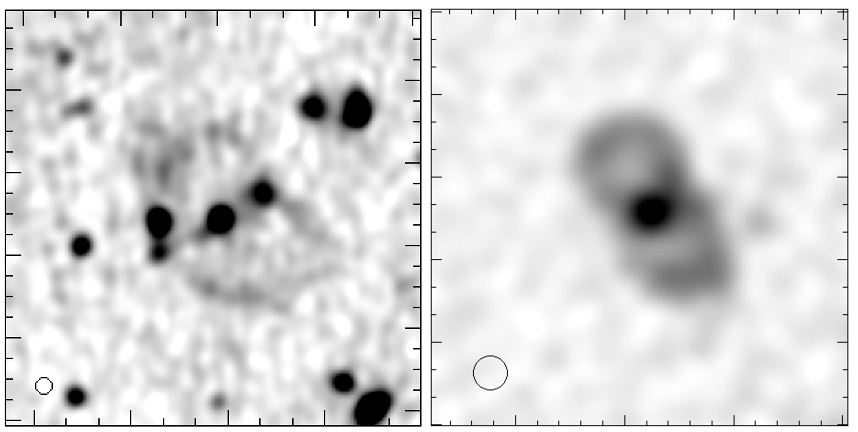}
\caption{ASKAP radio continuum images of two double ring systems with a massive elliptical galaxy in their intersect region:  
ORC~J0518--5105 (left) 
and ORC~J2207--5806 (right). 
The respective image sizes are 10\arcmin\ (left, 25\arcsec\ resolution) and 2.5\arcmin\ (right, 12.5\arcsec\ resolution).}
\label{fig:double-orcish}
\end{figure}

\subsection{Similar double rings in ASKAP}
High-resolution ASKAP radio continuum images from several large survey science projects are now available in CASDA. These include, for example, shallow images from the Rapid ASKAP Continuum Survey \citep[RACS,][]{McConnell2020, Duchesne2023} for the whole southern sky and up to northern declinations of $\sim$40\degr\ (15min. integration time per field), and deep images from the on-going EMU project \citep{Hopkins2025} for $\sim$30\% of the southern sky (10h integration time per field). While searching for ORCs we found several double ring systems (Koribalski et al., in prep.), two of which are shown in Figure~\ref{fig:double-orcish}. ORC~J2207--5806 is an unusual double ring source found in the EMU-PS \citep{EMU-PS}, and also noted by Norris, Yew et al. (2025, their Figure~14). The elliptical host galaxy has a redshift of \zph\ = $0.578 \pm 0.050$. The radio extent of 1.2\arcmin\ corresponds to 470~kpc. 
ORC~J0518--5105 is another stunning double ring, with a size of $\sim$4\arcmin. At the host galaxy redshift of \zph\ = $0.113 \pm 0.010$ this corresponds to $\sim$500~kpc. 

\subsection{Proposed ORC formation mechanism}

The original ORC discovery papers \citep{Norris2021, Koribalski2021} presented possible formation scenarios, which are re-examined and updated with every new ORC detection \citep[most recently, ][]{Koribalski-Veronica2024, Norris2025}. 
Numerical simulations of possible ORC formation mechanisms are essential to progress in this new field. \citet{Dolag2023} propose outwards moving merger shells as a likely mechanism to form ORCs with a wide range of radio morphologies, while \citet[][see Section~5.4]{Nolting2023} look into precessing jets and \citet{Shabala2024} model the collision of a remnant lobe with an external shock. 

A key feature of ORC~J1841--6547 are its intersecting double radio shells reminiscent of remnant lobes where only the outer "skin" (cocoon) remains. While faint, they look like two limb-brightened bubbles. Near circular double radio lobes like Fornax~A are not uncommon, but generally filled with diffuse gas and rarely edge-brightened. There is some evidence that radio lobes may be hollow with backflow along the jet-forged cocoon. 

\citet{Dolag2023} found outwards moving merger shells resembling ORCs in their high-resolution cosmological simulations of the evolution of massive elliptical galaxies and their CGM. \citet{Koribalski2024-Physalis} further expanded on this work, based on radio and X-ray observations of the nearby Physalis radio shell system, where merger shocks reignite remnant plasma within the group. ORC~J1841--6547 may be a case where merger shocks expanding inside the cocoon of remnant radio lobes, here nearly perpendicular to the line of sight, result in the edge-brightened double shell morphology. 

Double-double radio galaxies where re-started AGN jets propagate inside cocoons left behind by the initial jets, are used to infer the properties of the medium inside the cocoon \citep[e.g.,][]{Kaiser2000, yates2018}. Similarly, outwards moving merger shocks interact with the medium inside the cocoons, before becoming detectable as ORCs.


\section{Conclusions}
Our discovery of the double-shell system ORC J1841--6547 in deep ASKAP 944~MHz images adds diversity to the class of ORCs and prompted us to re-examine the proposed formation scenarios. The radio morphology of ORC J1841--6547 resembles that of two partial, intersecting shells with a massive elliptical galaxy, likely the system's host, in the centre of the intersect region. While the system somewhat resembles a fat double-lobed radio galaxy, only the outlines of lobes or bubbles are seen, lacking the typical diffuse emission filling. Instead, we detect edge-brightened emission in the form of radio shells, not unlike radio relics in the outskirts of galaxy clusters \citep{vanWeeren2019, Mandal2020}. The system spans 7\arcmin, ie., at least $\sim$1~Mpc at the host galaxy redshift (see Table~2). Each of the two shells/bubbles has a diameter of at least 540~kpc, similar to the size of ORC~1. \\ 

We propose that ORCs are radio relics around galaxy groups formed by outwards moving merger shocks during the evolution and growth of the central elliptical galaxy. Such relics are very rare and have not previously been noted in galaxy groups, while in galaxy clusters the occurrence of single or double radio relics in their outskirts is well established. Similar to clusters, the morphologies of radio relics around groups vary and depend on the properties of the IGM surrounding the central host galaxy. The ORC host galaxies have stellar masses of $\gtrsim$10$^{11}$\Msun\ and are notable as the brightest group galaxy (BGG). Our proposed formation scenario involves outwards moving merger shocks \citep{Dolag2023} to re-energise the magnetised plasma from fading radio lobes 
via (a) re-acceleration of pre-existing fossil CR electrons and (b) in situ acceleration by an ensemble of shocks with different Mach numbers formed in the turbulent ICM. We expect the radio shells / relics to be highly polarised, which can be confirmed by deep radio observations. Deep X-ray and spectroscopic optical observations should be invaluable to reveal (possibly strongly disturbed) thermal gas content of the system as well as possible signatures of the dense warm gas and tidal distortions of the stellar bodies.



\vspace{-0.3cm}

\section*{Acknowledgements}

We thank the ASKAP team for their dedicated and continuing work on creating such a powerful survey telescope, together with a robust data processing pipeline and public archive. A big thank you also to Ian Heywood who processed the MeerKAT 1.3~GHz data of ORC~5, which we show in Figure~7 (right). \\

IK acknowledges support by the COMPLEX project from the European Research Council (ERC) under the European Union’s Horizon 2020 research and innovation program grant agreement ERC-2019-AdG 882679. LMB is supported by NASA through grant 80NSSC24K0173. \\

This scientific work uses data obtained from Inyarrimanha Ilgari Bundara / the Murchison Radio-astronomy Observatory. We acknowledge the Wajarri Yamaji People as the Traditional Owners and native title holders of the Observatory site. CSIRO’s ASKAP radio telescope is part of the Australia Telescope National Facility (https://ror.org/05qajvd42). Operation of ASKAP is funded by the Australian Government with support from the National Collaborative Research Infrastructure Strategy. ASKAP uses the resources of the Pawsey Supercomputing Research Centre. Establishment of ASKAP, Inyarrimanha Ilgari Bundara, the CSIRO Murchison Radio-astronomy Observatory and the Pawsey Supercomputing Research Centre are initiatives of the Australian Government, with support from the Government of Western Australia and the Science and Industry Endowment Fund. \\

This project used data obtained with the Dark Energy Camera (DECam), which was constructed by the Dark Energy Survey (DES) collaboration. Funding for the DES Projects has been provided by the U.S. Department of Energy, the U.S. National Science Foundation, the Ministry of Science and Education of Spain, the Science and Technology Facilities Council of the United Kingdom, the Higher Education Funding Council for England, the National Center for Supercomputing Applications at the University of Illinois at Urbana-Champaign, the Kavli Institute of Cosmological Physics at the University of Chicago, Center for Cosmology and Astro-Particle Physics at the Ohio State University, the Mitchell Institute for Fundamental Physics and Astronomy at Texas A\&M University, Financiadora de Estudos e Projetos, Fundacao Carlos Chagas Filho de Amparo, Financiadora de Estudos e Projetos, Fundacao Carlos Chagas Filho de Amparo a Pesquisa do Estado do Rio de Janeiro, Conselho Nacional de Desenvolvimento Cientifico e Tecnologico and the Ministerio da Ciencia, Tecnologia e Inovacao, the Deutsche Forschungsgemeinschaft and the Collaborating Institutions in the Dark Energy Survey. The Collaborating Institutions are Argonne National Laboratory, the University of California at Santa Cruz, the University of Cambridge, Centro de Investigaciones Energeticas, Medioambientales y Tecnologicas-Madrid, the University of Chicago, University College London, the DES-Brazil Consortium, the University of Edinburgh, the Eidgenossische Technische Hochschule (ETH) Zurich, Fermi National Accelerator Laboratory, the University of Illinois at Urbana-Champaign, the Institut de Ciencies de l’Espai (IEEC/CSIC), the Institut de Fisica d’Altes Energies, Lawrence Berkeley National Laboratory, the Ludwig Maximilians Universitat Munchen and the associated Excellence Cluster Universe, the University of Michigan, NSF’s NOIRLab, the University of Nottingham, the Ohio State University, the University of Pennsylvania, the University of Portsmouth, SLAC National Accelerator Laboratory, Stanford University, the University of Sussex, and Texas A\&M University.

BASS is a key project of the Telescope Access Program (TAP), which has been funded by the National Astronomical Observatories of China, the Chinese Academy of Sciences (the Strategic Priority Research Program “The Emergence of Cosmological Structures” Grant \# XDB09000000), and the Special Fund for Astronomy from the Ministry of Finance. The BASS is also supported by the External Cooperation Program of Chinese Academy of Sciences (Grant \# 114A11KYSB20160057), and Chinese National Natural Science Foundation (Grant \# 12120101003, \# 11433005).

The Legacy Survey team makes use of data products from the Near-Earth Object Wide-field Infrared Survey Explorer (NEOWISE), which is a project of the Jet Propulsion Laboratory/California Institute of Technology. NEOWISE is funded by the National Aeronautics and Space Administration.

The Legacy Surveys imaging of the DESI footprint is supported by the Director, Office of Science, Office of High Energy Physics of the U.S. Department of Energy under Contract No. DE-AC02-05CH1123, by the National Energy Research Scientific Computing Center, a DOE Office of Science User Facility under the same contract; and by the U.S. National Science Foundation, Division of Astronomical Sciences under Contract No. AST-0950945 to NOAO.



\vspace{-0.3cm}

\section*{Data availability} 

The ASKAP data products used in this article are available through CASDA. Additional data processing and analysis was conducted using the {\sc miriad} software\footnote{https://www.atnf.csiro.au/computing/software/miriad/} and the Karma visualisation\footnote{https://www.atnf.csiro.au/computing/software/karma/} packages. DESI images were obtained with the Legacy Survey SkyViewer\footnote{https://www.legacysurvey.org/viewer/}.


\printendnotes

\bibliography{ORC6-pasa}

\end{document}